\documentclass[aps,pra,twocolumn,a4paper,groupedaddress,showpacs,floatfix,10pt]{revtex4-1}
\usepackage{graphicx,amsmath,amssymb,amsfonts,dsfont}
\newcommand{\mf}[1]{\boldsymbol{#1}}
\newcommand{\ket}[1]{\ensuremath{|#1\rangle}}
\newcommand{\mc}[1]{\ensuremath{\mathcal{#1}}}
\newcommand{\bra}[1]{\ensuremath{\langle #1 |}}

\usepackage{color}

\begin{document}

\title{Few-Body Bound States of Dipole-Dipole Interacting Rydberg Atoms}

\author{Martin Kiffner${}^{1,2}$}
\author{Mingxia Huo${}^{1,2}$}
\author{Wenhui Li${}^{1,3}$}
\author{Dieter Jaksch${}^{2,1}$}

\affiliation{Centre for Quantum Technologies, National University of Singapore,
3 Science Drive 2, Singapore 117543${}^1$}
\affiliation{Clarendon Laboratory, University of Oxford, Parks Road, Oxford OX1
3PU, United Kingdom${}^2$}
\affiliation{Department of Physics, National University of Singapore, 117542,
Singapore${}^3$}


%31.50.-x Potential energy surfaces

%32.80.Ee Rydberg states

%82.20.Rp State to state energy transfer

\begin{abstract}
We show that the resonant dipole-dipole interaction can give rise to bound states between 
two and three  Rydberg atoms with non-overlapping electron clouds. 
The dimer and trimer states arise from avoided level crossings between states converging to 
different fine structure manifolds in the limit of separated atoms. 
We analyze the angular dependence of the potential wells, characterize the quantum dynamics in these 
potentials and discuss methods for their production and detection. Typical distances between 
the atoms are of the order of several micrometers which can be  resolved  in state-of-the-art experiments. 
The potential depths and typical oscillation 
frequencies are about one order of magnitude larger as compared to the dimer and trimer states  
investigated in [PRA \textbf{86} 031401(R) (2012)] and [PRL \textbf{111} 233003 (2014)], respectively. 
We find that the dimer and trimer molecules can be aligned with respect to the axis of a weak electric field. 
\end{abstract}

\maketitle

\section{Introduction \label{intro}}
Rydberg atoms~\cite{gallagher:ryd} are 
atoms where at least one electron is in a highly excited state. 
State-of-the-art experiments offer unprecedented 
control over the position of cold and ultracold 
Rydberg atoms and allow one to prepare them in specific internal quantum states. 
These features combined with the exaggerated properties of Rydberg atoms make them 
ideally suited for investigating the quantum physics of few-body interactions. 
For example, the interaction of a single Rydberg electron with ground state atoms 
was investigated in~\cite{greene:00,bendkowsky:09,li:11,tallant:12,liu:09,balewski:13,krupp:14}. 
More specifically,   a single ground state atom interacting with a Rydberg electron 
gives rise to so-called trilobite molecules~\cite{greene:00,bendkowsky:09} that 
are several orders of magnitude larger than conventional molecules. In addition, 
these molecules can possess giant permanent electric dipole moments~\cite{li:11,tallant:12} 
and can be aligned by external magnetic fields~\cite{krupp:14}. 
It was shown that the interaction of one Rydberg electron with two ground state atoms 
can give rise to trimer states~\cite{liu:09}, and the interaction of a single Rydberg 
electron with a Bose-Einstein condensate was investigated in~\cite{balewski:13}.

A second example is given by the theoretically well-understood and tunable  
dipole-dipole (DD) interaction. Recently, the direct measurement 
of the van der Waals interaction between two Rydberg atoms was achieved~\cite{beguin:13}, 
and excellent agreement between theory and experiment was found. 
The DD interaction between Rydberg atoms is at the heart of  various exceptional phenomena in 
quantum optics~\cite{pritchard:12} and quantum information science~\cite{saffman:10}. 
Examples are given by the Rydberg blockade effect~\cite{jaksch:00,lukin:01,urban:09,gaetan:09}, 
the realization of quantum gates and entanglement~\cite{wilk:10,isenhower:10} and 
DD-induced artificial gauge fields acting on the relative motion of two Rydberg atoms were 
predicted in~\cite{zygelman:12,kiffner:13,kiffner:13b}. 
Moreover, several schemes have been developed where the DD interaction between Rydberg atoms 
with non-overlapping electron clouds gives rise to 
giant 
molecules~\cite{boisseau:02,schwettmann:06,schwettmann:07,overstreet:09,samboy:11,samboy:11b,samboy:13,kiffner:12,kiffner:13l}. 
Interatomic spacings in these so-called macrodimers typically exceed $1\,\mu\text{m}$ and thus 
their positions become  experimentally resolvable~\cite{schauss:12,schwarzkopf:11}. 
The binding mechanism of the macrodimer scheme proposed in~\cite{schwettmann:07} and 
observed in~\cite{overstreet:09} can be explained in terms of avoided crossings 
between Stark shifted states converging to van der Waals shifted two-atom states for large atomic separations. 
On the contrary, the dimer and trimer  states in~\cite{kiffner:12,kiffner:13l} arise 
from avoided crossings between Stark shifted states within a small manifold of near-resonantly coupled states. 
The resonant character of the DD interaction between those states reduces the number of relevant atomic levels.   
This feature leads to insightful and transparent physics and  allows one to account for the anisotropic 
nature of DD induced trapping potentials in the presence of external electric fields.

Here we investigate bound dimer and trimer states in DD interacting Rydberg atoms 
based on the Rydberg level scheme shown in Fig.~\ref{fig1}(a). This level scheme is 
an extension of the Rydberg macrodimer proposal in~\cite{kiffner:12} because it contains the  $np_{1/2}$ states in 
addition to the $np_{3/2}$ states. The purpose of the present work is twofold. First, we investigate 
the influence of the $np_{1/2}$ states on previous results~\cite{kiffner:12,kiffner:13l} and show that 
the qualitative results remain unchanged. Second, we find 
that the presence of the $np_{1/2}$ states gives rise to a novel type of  dimer and trimer states. These 
states are more deeply bound than the states reported in~\cite{kiffner:12,kiffner:13l} and do not 
require an external electric field. 
They can be explained in terms of avoided crossings between Rydberg states that are separated by 
the fine structure interval in the non-interacting limit of large atomic separations. 
We show that the dimer and trimer states arising from the presence of the $np_{1/2}$ states  
can be efficiently excited via microwave fields and find that the quantum dynamics in these wells 
is at least one order of magnitude faster as compared to previous results~\cite{kiffner:12,kiffner:13l}. 
A weak electric field breaks the spherical symmetry of the system and allows one to align the dimer and 
trimer configurations with respect to the field axis. 

This paper is organized as follows. We focus on two-atom bound states in Sec.~\ref{twobound} and 
discuss novel dimer states arising from the inclusion of the  $np_{1/2}$ states. 
We identify the physical origin of the binding mechanism and investigate the influence of the $np_{1/2}$ states 
on the bound states reported in~\cite{kiffner:12}. 
In Sec.~\ref{threebound} we analyze three-atom bound states  induced by  the mechanism explained 
in Sec.~\ref{twobound} and study the impact of the $np_{1/2}$ states on 
the trimer potential wells reported in~\cite{kiffner:13l}. 
The conclusion of our work is presented  in Sec.~\ref{conclusion}.
\section{Two-atom bound states \label{twobound}}
\begin{figure}[t!]
\includegraphics[width=\linewidth]{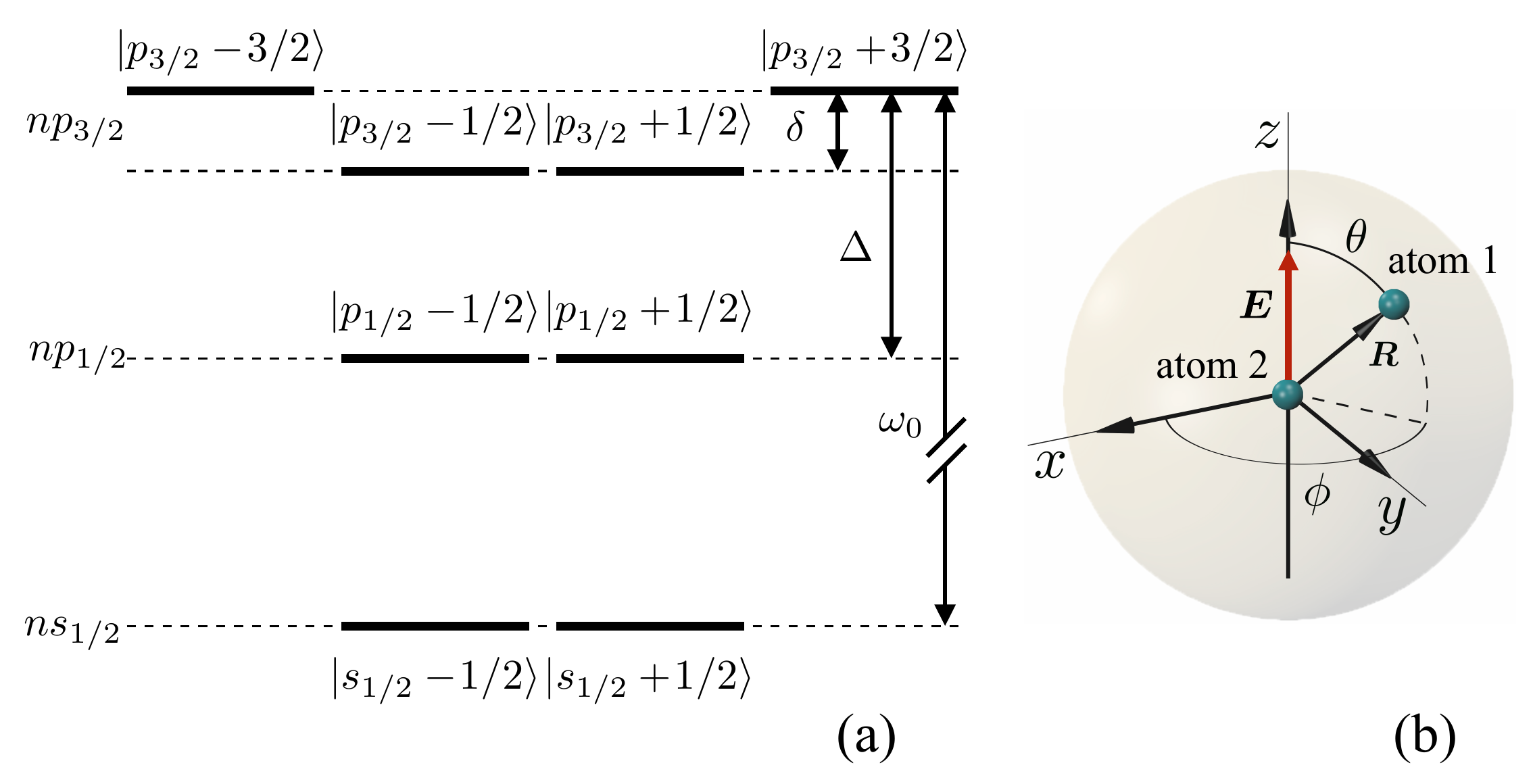}
\caption{\label{fig1}
(Color online) (a) Level structure of a single Rydberg atom including 
the $ns_{1/2}$, $np_{1/2}$ and $np_{3/2}$ Zeeman manifolds. The
fine structure interval $\Delta$ and the Stark splitting $\delta$ 
are defined  as $\hbar\Delta=E_{p_{1/2}\pm1/2}-E_{p_{3/2}\pm3/2}<0$ and 
$\hbar\delta=E_{p_{3/2}\pm1/2}-E_{p_{3/2}\pm3/2}<0$, respectively. 
Note that both parameters $\Delta$ and $\delta$ are negative.
(b) Two DD interacting atoms with relative position $\mf{R}$. 
An  electric field $\mf E$ is applied along the $z$ direction.
}
\end{figure}
We investigate two-atom bound states generated by the DD interaction for Rydberg
atoms with the internal level structure shown in Fig.~\ref{fig1}(a) and placed
into an external electric field $\mf{E}$. 
Deeply bound molecular states arise because of the inclusion of 
the $np_{1/2}$ manifold. Their size is determined by the length scale 
$r_0$ which is the distance between two Rydberg atoms where the DD interaction 
equals the fine structure interval $\hbar |\Delta|$. The effect of the 
$np_{1/2}$ states on the bound dimer states  investigated in
\cite{kiffner:12}, which appear at larger atomic distances $R_0$ where $\hbar
|\delta|$ equals the strength of the DD interaction, is negligible.

In Sec.~\ref{disystem} we describe the two-atom interacting system and
define the relevant length and energy scales. This is followed by a qualitative
discussion of the binding mechanism in Sec.~\ref{dimechanism}. In
Sec.~\ref{dipotential} we discuss the properties of the potential surfaces
leading to bound molecular states and the quantum dynamics of the relative atomic
motion in these potentials. We conclude the section by describing possible
methods for preparing and detecting the bound molecular states.
\subsection{The system \label{disystem}}
We first consider a single alkali atom with Rydberg energy level structure
$\ket{l_jm}$ and principal quantum number $n\gg 1$ as shown in
Fig.~\ref{fig1}(a). In our standard notation $l$ labels the orbital angular
momentum of the Rydberg excited valence electron and $j$ its total angular
momentum. The projection of the electron's angular momentum onto the $z$-axis is
denoted by $m$. The levels $\ket{s_{1/2}\pm 1/2}$ are separated from the
$\ket{p_{3/2}\pm 3/2}$ levels by energy $\hbar \omega_0$. An external electric
field $\mf{E}$ along the $z$-direction induces a Stark shift $\delta$ between
states $\ket{p_{3/2} \pm 1/2}$ and $\ket{p_{3/2} \pm 3/2}$. So far this level
structure is identical to that considered in \cite{kiffner:12,kiffner:13l}. In
contrast to this previous work we here also include the $np_{1/2}$ states into
our considerations. These are separated from the $np_{3/2}$ states by the fine
structure splitting $\Delta$ 
which will typically be much larger than the electric field induced Stark shift
$\delta$. Specifically, in our calculations we will assume $\Delta/\delta =
100$. This choice corresponds to weak electric fields of the order of $1\,\text{V}/\text{cm}$~\cite{kiffner:13l} 
such that the mixing of states with opposite parity is negligible~\cite{kiffner:12}. 
For Rubidium atoms with $n\ge 30$ we find~\cite{li:03} $\omega_0\approx 40\,|\Delta|$, and 
thus the energy $\hbar\omega_0$ of the $ns-np$ interval is much larger than any energies of interest in this work. 
The internal atomic Rydberg levels in Fig.~\ref{fig1}(a) are described by the Hamiltonian
$H_{\text{A}}$ whose explicit form is given in Appendix~\ref{appena}.

The Hamiltonian describing the relative motion and internal degrees of freedom
of two DD interacting atoms is given by
\begin{align}
H=\frac{\mf{\hat{p}}^{2}}{2\mu} + H_{\text{int}}. \label{Ht}
\end{align}
Here $\mf{\hat{p}}$ is the relative momentum operator canonically conjugate to
the distance operator between the atoms $\mf{\hat R}$. The reduced mass of the
two-atom system is denoted by $\mu$. The second term $H_{\text{int}}$ describes
the internal dynamics of the two atoms and can be expressed as
\begin{equation}
H_{\text{int}} = \sum\limits_{\alpha=1}^{2} H_{\text{A}}^{(\alpha)} +V_{12}(\mf{\hat R}),
\label{hint}
\end{equation}
where $H_{\text{A}}^{(\alpha)}$ describes the energy levels of atom $\alpha$ and
$V_{12}(\mf{R})$ is the DD
interaction between the two atoms at separation $\mf R$ as detailed in
Appendix~\ref{appena}. Note the lack of an external trapping potential for the
Rydberg energy levels. 

In this work we  use the Born-Oppenheimer approximation to obtain the adiabatic energy surfaces 
for the relative atomic motion.
Following the approach in~\cite{kiffner:12} we diagonalize the Hamiltonian $H_{\text{int}}$ replacing $\mf{
\hat R} \rightarrow \mf{R}$. This calculation (for details see
Appendix~\ref{appena}) yields eigenstates of the internal degrees of freedom
which parametrically depend on the interatomic separation $\mf R$. 
For all potential surfaces describing bound states we have verified 
via semi-classical calculations~\cite{kiffner:13l} that 
the atomic motion on potential wells remains adiabatic and hence this approximation is 
justified. The reason is that the potential surfaces describing bound states 
are energetically well separated from other curves, and the considered velocity of the 
cold atoms in the wells is sufficiently small. 

Before quantitatively investigating these energy surfaces we describe the
physical mechanism that leads to potential minima and hence bound molecular
states of the two atoms. The length scale at which these bound states occur is
determined by the distance $r_0$ between the atoms where the DD interaction
strength equals the energy separation between the $np_{3/2}$ and the $np_{1/2}$
levels. As shown in Appendix~\ref{appena} this value is given by 
\begin{align}
r_0=\left[\frac{|\mathcal{D}|^2}{4 \pi \varepsilon_0\hbar |\Delta|}\right]^{1/3}, 
\label{r0}
\end{align}
where $\mathcal{D}$ is a reduced dipole matrix element defined in Appendix~\ref{appena} 
and $\varepsilon_0$ is the dielectric constant. We note that this definition
is different from $R_0$ in previous work \cite{kiffner:12} where the
characteristic molecule size was determined by $\delta$ instead of $\Delta$.
While $\Delta$ will typically be two orders of magnitude larger than $\delta$
the value of $r_0$ reduces only with the cube root of this frequency ratio hence
still giving molecules with sizes on the order of microns. For instance, in the
case of Rb atoms with $n=40$, the splitting is $\Delta \simeq 2 \pi \times 1$GHz,
which yields $r_{0} \approx 1\mu$m.
\subsection{Binding mechanism \label{dimechanism}}
In this section we provide a simple explanation for the formation of bound
states in
DD interacting Rydberg atoms. We describe the physical mechanism by means of the
two-atom
system shown in Fig.~\ref{fig1}, but the general idea does
also apply to
the dimer and trimer states investigated in ~\cite{kiffner:12}
and~\cite{kiffner:13l}, respectively.

\begin{figure}[t!]
\begin{center}
\includegraphics[width=8.5cm]{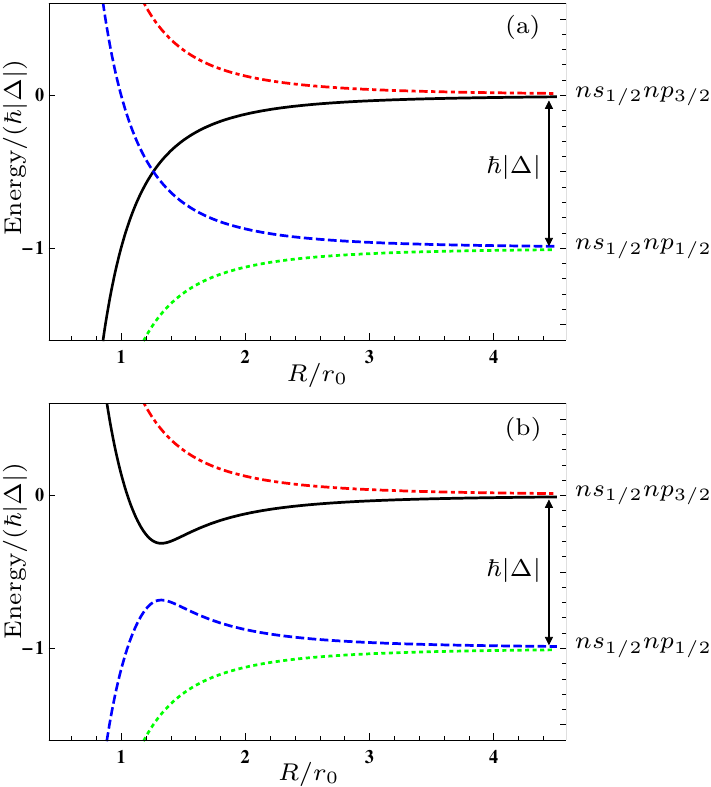}
\end{center}
\caption{\label{fig2}
(Color online) Simplified level structure of the DD coupled $nsnp$ states
ignoring all Zeeman sublevels of the level scheme in Fig.~\ref{fig1}(a).
(a) Position-dependent energies within the  $ns_{1/2}np_{3/2}$ ($ns_{1/2}np_{1/2}$) manifold
are shown by the black solid and red dot-dashed
(blue dashed and green dotted) lines. The cross-coupling $\Omega_{\text{cc}}$ in
Eq.~(\ref{cc}) between different manifolds is assumed to be zero.
(b) Same as in (a) but with  $\Omega_{\text{cc}}\not=0$ resulting in an avoided
level crossing between the black solid  and blue dashed lines.
}
\end{figure}

We consider the subspace of $nsnp$ states that are directly coupled via the DD
interaction.
In order to describe the general level structure of DD-coupled two-atom states
we ignore the Zeeman sublevels and any Stark shifts between them. In this case,
each atom reduces to a three-level system with
states $\ket{s_{1/2}}$, $\ket{p_{1/2}}$ and  $\ket{p_{3/2}}$. The fine structure
splitting $\Delta$ gives rise to two different
energy asymptotes for large values of $R$. They correspond to the  $ns_{1/2}np_{3/2}$
and $ns_{1/2}np_{1/2}$ manifolds
differing in energy by $\hbar|\Delta|$ for $R\rightarrow\infty$, see
Fig.~\ref{fig2}(a).
The $ns_{1/2}np_{3/2}$ manifold consists of the states $\ket{s_{1/2},p_{3/2}}$ and
$\ket{p_{3/2},s_{1/2}}$ that are resonantly coupled
by the DD interaction. Diagonalization of the DD interaction within the
$ns_{1/2}np_{3/2}$ manifold leads to the
potential curves shown by the  black solid and red dot-dashed lines in
Fig.~\ref{fig2}(a).
The energy splitting between them scales as $1/R^3$.
An equivalent analysis applies to $ns_{1/2}np_{1/2}$ manifold comprising the states
$\ket{s_{1/2},p_{1/2}}$ and $\ket{p_{1/2},s_{1/2}}$.
The DD interaction between them results in position-dependent energies that are
shown by the blue dashed  and green dotted lines in Fig.~\ref{fig2}(a).
%
%%%%%%%%%%%%%%%%%%%%%%
%%%%%%%%%%%%%%%%%%%%
%

%
So far we have ignored the DD coupling between states belonging to different
manifolds. This is the reason why the blue dashed and black solid lines
in Fig.~\ref{fig2}(a) cross. In general, the DD interaction will couple states
in the $ns_{1/2}np_{3/2}$ manifold with those in $ns_{1/2}np_{1/2}$, i.e.,
\begin{align}
\Omega_{\text{cc}} = |\bra{s_{1/2},p_{3/2}}V_{12}\ket{p_{1/2},s_{1/2}}|\not=0 .
\label{cc}
\end{align}
The DD coupling between the blue dashed  and black solid lines in
Fig.~\ref{fig2}(b) is van der Waals like for large separations $R\gg r_0$ and
turns into a resonant DD interaction near  $R\approx r_0$, where the
characteristic length scale $r_0$ is defined in Eq.~(\ref{r0}).
Most importantly, a non-zero cross-coupling $\Omega_{\text{cc}}$ results in an
avoided level crossing between the  black solid and
blue dashed  curves giving rise to a  potential minimum, see Fig.~\ref{fig2}(b).

In the following section~\ref{dipotential} we investigate the level structure of
the DD coupled $nsnp$ states taking into account the full
level scheme shown in Fig.~\ref{fig1}(a).
\subsection{Potential surfaces \label{dipotential}}
%
%
%%%%%%%%%%%%%%%%%%%%%%
\begin{figure}[t!]
\begin{center}
\includegraphics[width=8.5cm]{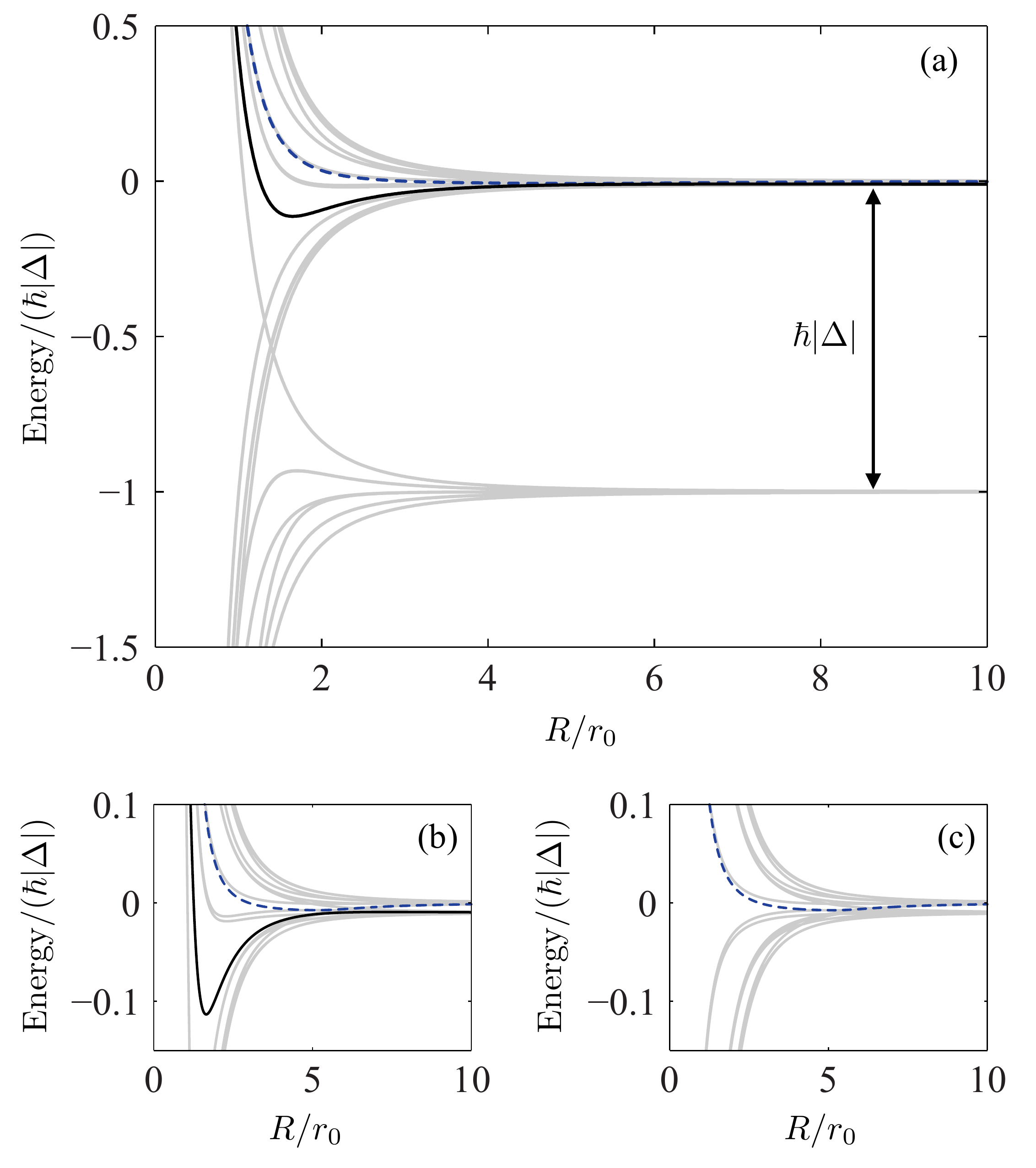}
\end{center}
\caption{\label{fig3}
(Color online) (a) All potential curves in the $nsnp$ manifold. 
 The black solid line shows the deepest trapping potential 
and the blue dashed line corresponds to the dimer state investigated in~\cite{kiffner:12}. 
(b)  Magnified view of potential curves converging to the $ns_{1/2}np_{3/2}$ asymptote. 
Note the changed  energy scale as compared to (a). 
(c)  All states in the nsnp manifold but without the $np_{1/2}$ states as in~\cite{kiffner:12}.
In (a)-(c), the parameters are $\Delta/\delta=100$ and $\theta=\pi/2$.
}
\end{figure}
%%%%%%%%%%%%%%%%%%%%%%%%%%%%%%%%
%
We consider the subspace of $nsnp$ states that are near-resonantly coupled by the DD interaction. 
Other two-atom states cause a negligible van der Waals shift for 
$R\ge r_0$ if their energy separation from the $nsnp$ manifold is large as compared to $\hbar |\Delta|$. 
Note that such a clear separation between resonantly and off-resonantly coupled states is  not possible in other 
schemes~\cite{boisseau:02,schwettmann:06,schwettmann:07,overstreet:09,samboy:11,samboy:11b,samboy:13} 
which are based on a large manifold of van der Waals coupled states. 
In Fig.~\ref{fig3}(a) we plot the 
$nsnp$ eigenstates as a function of atomic distance $R$ and angle $\theta =
\pi/2$. For large distances the manifolds of $ns_{1/2}np_{1/2}$ and $ns_{1/2}np_{3/2}$
states are seen to be separated by energy $\hbar |\Delta|$. As described above
when the DD interaction becomes comparable to this energy at around $r_0$ it
leads to avoided crossings between curves of the two manifolds giving rise to
potential wells. The deepest well with a minimum near $R_p\approx 1.7 r_0$ is shown as a black solid
 line in Fig.~\ref{fig3}(a). The well depth is approximately given by $0.1\hbar|\Delta|$ 
corresponding to trapping frequencies of the order of $100\,\text{MHz}$ for $n\approx 40$. 
Note that the influence of the Stark shift $|\delta|\ll|\Delta|$ on this potential well is small. 
More details and the molecular states bound by this potential
well will be analyzed in~\ref{diqd}.

In Fig.~\ref{fig3}(b) we show a magnified view of  the region investigated in~\cite{kiffner:12}. 
This is compared to the potential curve first described there which is plotted 
in Fig.~\ref{fig3}(c) ignoring the $ns_{1/2}np_{1/2}$ manifold. 
The minimum of the 
potential is located at approximately $R_0 = 100^{1/3} r_0 \approx 4.6 r_0$
where the DD coupling between the $ns_{1/2}np_{1/2}$ and $ns_{1/2}np_{3/2}$ manifolds is 
negligible as confirmed by comparison with Fig.~\ref{fig3}(b). 
We find that in the vicinity of $R_0$ the shape of the potential well, and hence
the properties of the corresponding molecular states, are nearly unchanged by
the presence of $np_{1/2}$ states. Only for smaller separations we obtain a
noticeable change of the curve which, however, does not affect the properties of
bound states discussed in \cite{kiffner:12}.
\subsubsection{Quantum dynamics \label{diqd}}
%
%%%%%%%%%%%%%%%%%%%%%%
\begin{figure}[t!]
\includegraphics[width=8cm]{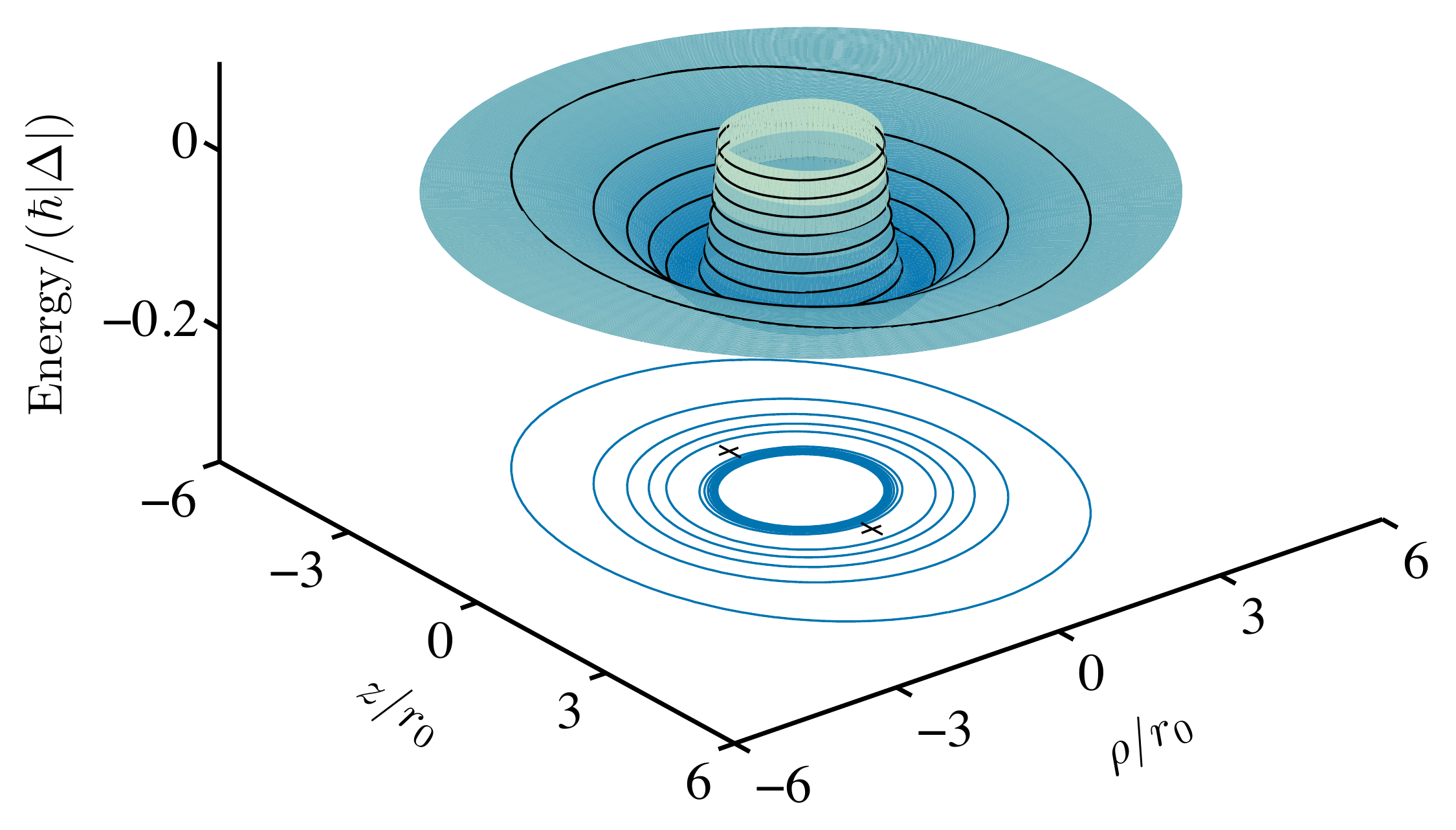}
\caption{\label{fig4}
(Color online) The  potential well indicated by the black solid line in Fig.~\ref{fig3}(a) in 
the $\rho-z$ plane for $\Delta/\delta=100$. The two absolute minima of the potential well 
along the $z$ axis are indicated by crosses in the contour projection of the potential. 
}
\end{figure}
%%%%%%%%%%%%%%%%%%%%
%
In Fig.~\ref{fig4} we show the energy surface of the deeply bound molecular 
potential of Fig.~\ref{fig3}(a) as a function of $\rho = \sqrt{x^2+y^2}$ and $z$. 
Due to the  electric field in $z$ direction and the associated Stark shift the potential well 
is not isotropic but the azimuthal symmetry is preserved. Since $|\delta|\ll|\Delta|$ 
the asymmetry of the potential well in Fig.~\ref{fig4} is not very pronounced. However, 
there are two absolute minima along the $z$ axis at $R_p\approx 1.7 r_0$. 
This allows one to align the molecule in the direction of the electric field. 
Expanding the potential around one of its minima to second order and 
diagonalizing the resulting Hessian matrix~\cite{kiffner:13l}   we obtain the vibrational
frequencies of the molecular motion $\omega_{1}= \omega_{2}\approx 0.07 \omega_{\text{vib}}$
and $\omega_{3} \approx 0.80 \omega_{\text{vib}}$ where
\begin{equation}
\omega_{\text{vib}} = \sqrt{\frac{\hbar|\Delta|}{\mu r_{0}^{2}}}.
\label{ovib}
\end{equation}
The typical vibrational frequency is about $100^{5/6}\approx 46.4$ larger than the 
corresponding frequency in~\cite{kiffner:12,kiffner:13l} where the characteristic energy scale 
was given by the  Stark shift $\delta$ instead of $\Delta$. It follows that the quantum dynamics 
in the new dimer states is roughly one order of magnitude faster. 
For $\delta=0$ the potential surfaces become isotropic and only one vibrational mode remains. The two 
other modes are rotational with $E_{\text{rot}}=\ell (\ell +1) \hbar ^{2} / 2 I$, where
$\ell$ is the non-negative integer rotational quantum number and $I \simeq \mu
r_0^2$ denotes the typical moment of inertia.

The lifetime of the molecules is limited by the lifetime of Rydberg atoms due to
spontaneous decay \cite{kiffner:12}. At a temperature of $300\mathrm{\mu K}$ we
estimate a decay rate of approximately $25$kHz for Rydberg states in ${}^{85}$Rb
with $n=40$ \cite{beterov:09}. For the same parameters we find $\omega_1 =\omega_2
\approx 2 \pi \times 36$kHz, $\omega_3 \approx 2 \pi \times 419$kHz and $\hbar /I \approx 2\pi
\times 0.1$ $\text{kHz}$ indicating that molecular vibrations can be resolved
within the life time of the molecules while rotational excitations will not be
observable. The  equilibrium distance between the atoms for $n=40$ is given by $R_p\approx 1.7\mu\text{m}$. 
This value increases to $R_p\approx 4.0\mu\text{m}$ for $n=60$ which is experimentally 
attainable~\cite{schauss:12,schwarzkopf:11,beguin:13}. 
\subsubsection{Preparation and detection \label{dipd}}
\begin{figure}[t!]
\begin{center}
\includegraphics[width=7cm]{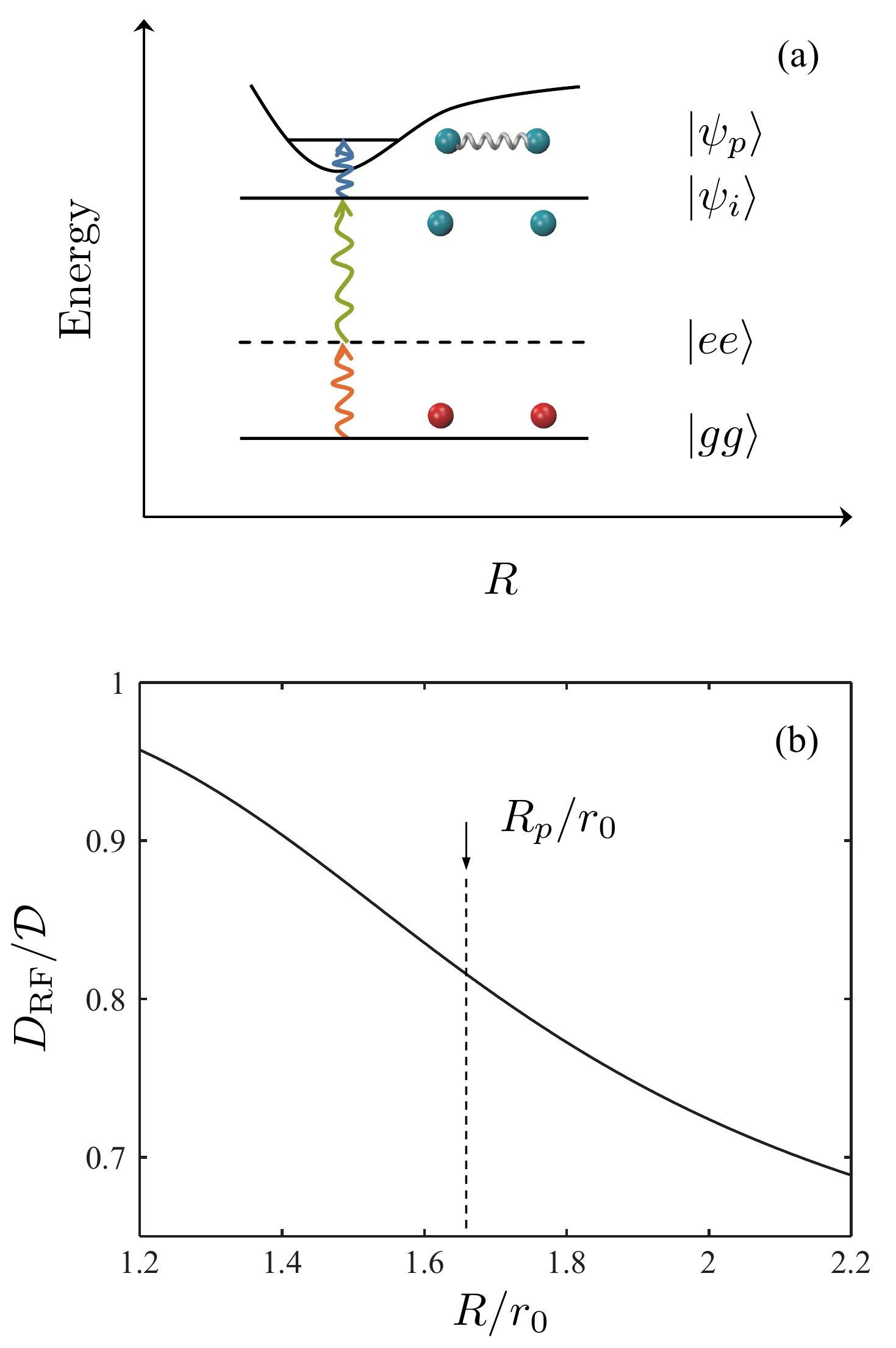}
\end{center}
\caption{\label{fig5}
(Color online) 
(a) Schematic illustration of the  excitation scheme. A
two-photon transition couples the  ground  state $\ket{gg}$ to a
two-atom Rydberg state $\ket{\psi_i}$ via an intermediate  state $\ket{ee}$. The atoms are excited by
microwave fields from the $\ket{\psi_i}$ to a bound state $\ket{\psi_p}$. 
(b) Transition dipole matrix element
between a particular bound state $\ket{\psi_p}$ [see text for details]
and the initial state $\ket{\psi_i}$ in Eq.~(\ref{gstateM})
for $\sigma^-$ polarized light.
The parameters are $\Delta/\delta=100$, $\theta=0$ and $\mc{D}$ is defined in 
Eq.~(\ref{D}).
} \label{preparations}
\end{figure}

Recent experimental progress has enabled the trapping of individual ultracold
atoms separated by a controllable distance on the order of $r_0$ using
microscopic optical traps \cite{beguin:13}. We may hence assume the preparation
of Rydberg molecules to start from two ground state atoms separated by the size
of the targeted Rydberg molecule state $\ket{\psi_p}$. A two-photon excitation
\cite{li:05,park:11,park:11b} as schematically shown in Fig.~\ref{fig5}(a) can
then be used to promote the atoms from the ground state manifold $\ket{gg}$ via
an intermediate optically excited state $\ket{ee}$ to the state
\begin{align}
\ket{\psi_i} = \ket{s_{1/2}+1/2,s_{1/2}+1/2}
\label{gstateM}
\end{align}
in the $nsns$ manifold. The atoms in the $nsns$ manifold only weakly interact
via the van der Waals interaction leading to a  blockade radius much smaller
than their separation. Also, atoms excited to a Rydberg state will no longer be
trapped by the optical potential \cite{beguin:13}. This is followed by a
microwave excitation, also shown in Fig.~\ref{fig5}(a), to $\ket{\psi_p}$ in the
$nsnp$ manifold which is bound solely by the DD interaction.
%
%
%%%%%%%%%%%%%%%%%%%%%%
%%%%%%%%%%%%%%%%%%%%
%

For left-circularly polarized light $\vec{\mf{\epsilon}}_{1}$ the Rabi frequency
of the microwave transition to the bound state is proportional to
\begin{equation}
D_{\text{RF}}=|\bra{\psi_p}
\mf{\hat{d}}_{\text{2atom}}\cdot \vec{\mf{\epsilon}}_{1}\ket{\psi_i}|,
\label{dtwo}
\end{equation}
where
\begin{equation}
\mf{\hat{d}}_{\text{2atom}}=\mf{\hat{d}}^{(1)}+\mf{\hat{d}}^{(2)}\,.
\end{equation}
The dependence of $D_{\text{RF}}$ on the interatomic separation $R$ for
$\theta=0$ is shown in Fig.~\ref{fig5}(b). Since $D_{\text{RF}}$ is large at the
position of the potential minimum where the Franck-Condon factor associated with
the transition $\ket{\psi_i}\rightarrow\ket{\psi_p}$ is
maximal, the molecular dimer state can be excited efficiently via microwave
radiation.

The extreme sensitivity of Rydberg atoms to external electric fields can be used
for state selectively ionizing the atoms and then detecting the resulting
electrons and ions with high efficiency. Specifically, the bound state
$\ket{\psi_p}$ can be experimentally observed by first exciting it to a
higher-lying $nsn'd$ state, and then detecting the $n'd$ atom via state-selective field 
ionization \cite{li:05,park:11,park:11b}. 
\section{Three-atom bound states \label{threebound}}
\begin{figure}[t!]
\begin{center}
\includegraphics[width=8.5cm]{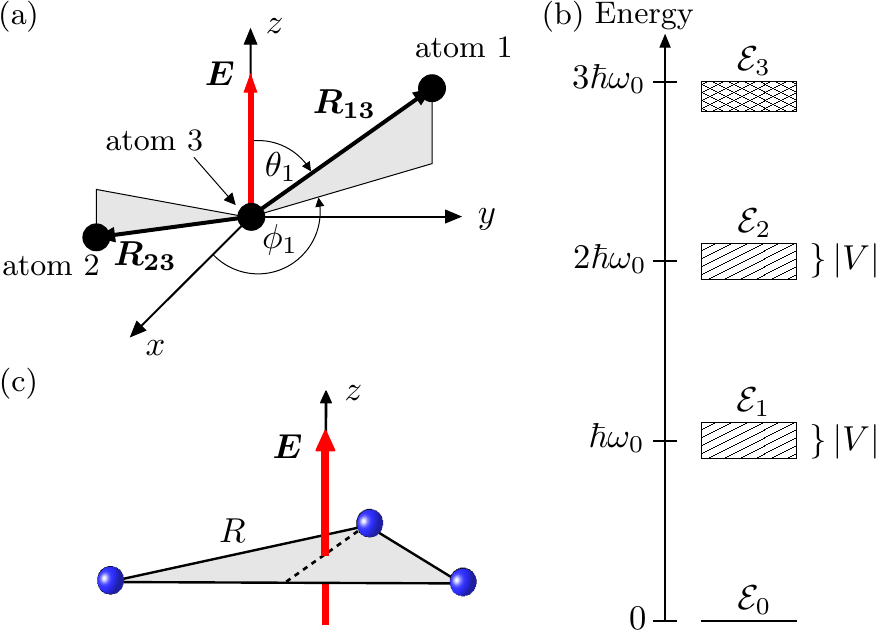}
\end{center}
\caption{\label{fig6}
(Color online) 
(a) System configuration of three DD interacting 
Rydberg atoms in an external electric field $\mf{E}$. 
The relative position vectors $\mf{R}_{\alpha 3}$  ($\alpha\in\{1,2\}$) are
expressed in terms of 
spherical coordinates $R_{\alpha 3}$, $\theta_{\alpha}$ and $\phi_{\alpha}$. 
The angles $\theta_{1}$ and $\phi_{1}$ are indicated in the figure, while 
$\theta_{2}$ and $\phi_{2}$ were omitted in order to keep the drawing concise. 
(b) Level structure of the three-atom state space. 
$\mc{E}_{i}$ contains all states where $i$ atoms are in an $np$ state 
and all other in an $ns$ state. States within  $\mc{E}_1$ and $\mc{E}_2$ 
are coupled by the DD interaction.
(c) Geometry where  the atoms form the vertices of 
an equilateral triangle with  edge length $R$ in the $x - y$ plane. 
The $z$ direction is distinguished by the external electric field $\mf{E}$. 
}
\end{figure}

Next we consider  three DD interacting Rydberg atoms placed in an external electric 
field $\mf{E}$, see Fig.~\ref{fig6}(a).  
This setup was introduced in~\cite{kiffner:13l}, where it was found that the DD
interaction can give rise to three-body bound states. 
These trimer states represent a genuine three-particle effect because they
cannot be explained by a pairwise binding of the atoms. 
They arise from a variation of the mechanism described in Sec.~\ref{dimechanism} where the 
fine structure splitting is replaced by Stark shifts between Zeeman states. This binding mechanism 
was shown in~\cite{kiffner:13l} to be fundamentally different from the Efimov 
systems in~\cite{kraemer:06,pollack:09,wang:11}, where a resonant two-body
interaction can be described 
by a single scattering length exceeding all physically relevant length
scales~\cite{efimov:70,braaten:06}. 
The aim of this section is to investigate DD induced trimer states in the system
shown in Fig.~\ref{fig6}(a) and for the 
level scheme in Fig.~\ref{fig1}(b). This extends the original study
in~\cite{kiffner:13l} where only the $ns_{1/2}$ and $np_{3/2}$ 
multiplets were considered. 
In Sec.~\ref{sys}, we briefly describe the system and divide the state space of
the three atoms in 
subspaces in order to facilitate the analysis. 
We find that the general mechanism for the formation of bound states described
in 
Sec.~\ref{dimechanism} gives rise to  more deeply bound trimer states as
compared to the states in~\cite{kiffner:13l} 
[see Sec.~\ref{compare}]. Furthermore, we investigate the  influence of the
$np_{1/2}$ states on the trimer states 
reported in~\cite{kiffner:13l}. 
In Sec.~\ref{new} we discuss in detail one trimer state in the $nsnsnp$ manifold
which only arises if the $np_{1/2}$ states are taken into account. 
The latter state  has a direct transition dipole moment with states in the
$nsnsns$ manifold which facilitates its preparation. 
%
%
%%%%%%%%%%%%%%%%%%%%%%
%%%%%%%%%%%%%%%%%%%% 
%
\subsection{The system \label{sys}}
The internal states of the three-atom  system in Fig.~\ref{fig6}(a) 
are determined by the Hamiltonian 
\begin{align}
H_{\text{int}} =  \sum\limits_{\alpha=1}^{3} H_{\text{A}}^{(\alpha)} 
+ V_{13}(\mf{\hat{R}}_{13}) + V_{23}(\mf{\hat{R}}_{23}) + V_{12}(\mf{\hat{R}}_{12}). 
\label{H}
\end{align}
Note that all DD interactions between the three atoms described by $H_{\text{int}}$ are pairwise interactions. 
Next we briefly recall some of the notation introduced in~\cite{kiffner:13l}. 
First, we express the two relative position vectors $\mf{R}_{\alpha 3}$ 
($\alpha\in\{1,2\}$) in terms of spherical coordinates,  
\begin{align}
\mf{R}_{\alpha 3} = R_{\alpha 3} (\sin\theta_{\alpha}
\cos\phi_{\alpha},\sin\theta_{\alpha} 
\sin\phi_{\alpha}, \cos\theta_{\alpha}). 
\end{align} 
It follows that the Born-Oppenheimer surfaces of the  Hamiltonian in
Eq.~(\ref{H}) can be 
characterized in terms of the five independent variables  
\begin{align}
\mf{v} = (R_{13},R_{23},\theta_1,\theta_2,\phi),
\label{vars}
\end{align}
where  $\phi=\phi_1-\phi_2$. Note that $\phi_1$ and $\phi_2$ are not independent
variables because the 
azimuthal symmetry of the system makes the energies independent of
$\phi_1+\phi_2$. 
Second, the three-atom states of the system can be  conveniently grouped in 
four  subspaces $\mc{E}_{i}$ ($i\in\{0,1,2,3\}$), where 
$\mc{E}_{i}$ contains all three-atom states with $i$ atoms in an $np$ state 
and $3-i$ atoms  in an $ns$ state. 
States in $\mc{E}_{i}$ are clustered in energy around $i\times \hbar \omega_0$
as shown in Fig.~\ref{fig6}(b). 
Due to their large energy separation we neglect any DD induced 
coupling between them  and diagonalize $H$ in Eq.~(\ref{H}) in each subspace
$\mc{E}_{i}$ independently. We have verified numerically that this is an excellent approximation for 
the parameter regime considered in the remaining part of this section.
Finally, we recall that the Hamiltonian in Eq.~(\ref{H})  is time-reversal
invariant~\cite{haake:qsc,kiffner:13l} 
and gives rise to Kramers degeneracy.  Every eigenvalue of $H_{\text{int}}$ in Eq.~(\ref{H})
is thus (at least) two-fold degenerate.
In the following Secs.~\ref{compare} and ~\ref{new} we investigate the level
structure of the  subspaces 
$\mc{E}_1$ and $\mc{E}_2$ and the trimer states within them. 
\subsection{Potential surfaces in $\mc{E}_2$ \label{compare}}
\begin{figure}[t!]
\begin{center}
\includegraphics[width=8.5cm]{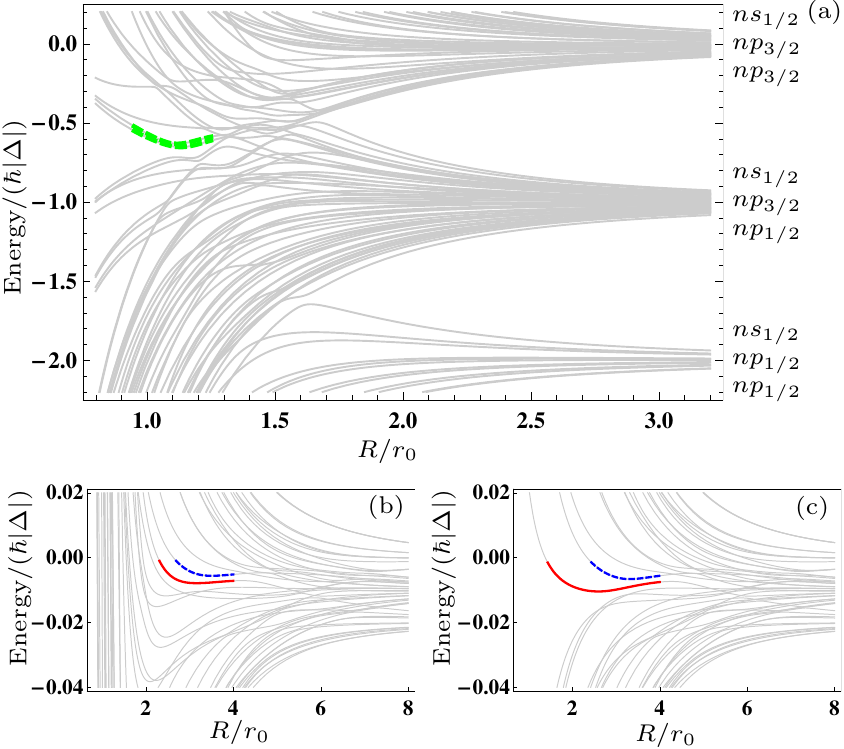}
\end{center}
\caption{\label{fig7}
%mathem
(Color online) 
(a) Potential curves within the manifold $\mc{E}_2$ for $\Delta/\delta=100$. 
(b) Magnified view of potential curves converging to the $ns_{1/2}np_{3/2}np_{3/2}$ asymptote. 
Note the change in energy and length scales as compared to (a). 
The  red solid and blue dashed curves correspond to the trimer configurations
discussed in~\cite{kiffner:13l}. 
(c) Same as in (b), but without the $p_{1/2}$ states as in~\cite{kiffner:13l}. 
}
\end{figure}

We consider the  geometrical setup shown in Fig.~\ref{fig6}(c)  and 
investigate three-body bound states in the subspace $\mc{E}_2$. All potential
curves in $\mc{E}_2$ are shown 
in Fig.~\ref{fig7}(a) as a function of the triangle edge length $R$. 
There are three different independent atom asymptotes at energies  $0$,
$\hbar\Delta$ and  $2\hbar\Delta$ for large values of $R$, 
corresponding to the $nsnp_{3/2}np_{3/2}$, $nsnp_{3/2}np_{1/2}$ and 
$nsnp_{1/2}np_{1/2}$ manifolds, respectively. 
Several avoided crossings between potential curves  are visible in
Fig.~\ref{fig7}(a), giving rise to 
potential wells according to the mechanism explained in Sec.~\ref{dimechanism}. 
Two wells are highlighted by the green  dashed 
lines in Fig.~\ref{fig7}(a) as an example.  
The minima occur in a region between $r_0$ and $1.7 r_0$ and are also present 
without an electric field and the associated Stark shift $\delta$ between Zeeman
sublevels in the $np_{3/2}$ manifold. 
We find that the influence of the Stark shift $\delta$ is negligible on the 
energy scale set by the fine structure splitting $\Delta$ 
and for the parameters chosen in Fig.~\ref{fig7}(a). 
Next we investigate the influence of the $np_{1/2}$ states on the trimer states
reported in~\cite{kiffner:13l}. In the latter system the depth of the potential wells 
are of the order of the Stark splitting $|\delta|$  which is typically much smaller than 
the fine structure interval $\hbar|\Delta|$. 
Figure~\ref{fig7}(b) shows a magnified view of some of the potential curves in
Fig.~\ref{fig7}(a) and on an energy scale that is comparable 
to $\hbar|\delta|$. 
A lot of the potential curves in Fig.~\ref{fig7}(b) have local minima, and the
trimer potentials investigated in~\cite{kiffner:13l} are represented by the 
red solid and blue dashed curves. 
The corresponding potential curves for the reduced level scheme considered
in~\cite{kiffner:13l} are shown in Fig.~~\ref{fig7}(c). 
A comparison of Figs.~~\ref{fig7}(b) and (c) shows that the $np_{1/2}$ states
shift the minima  of the potential curves  towards larger values. 
The equilibrium value of $R$ for 
the red solid (blue dashed) curve in Fig.~~\ref{fig7}(b) is 1.22 (1.06) times 
larger than for the case without $np_{1/2}$ states in Fig.~~\ref{fig7}(c). 
In addition, the presence of the $np_{1/2}$ states compresses the potential curves leading 
to larger oscillation frequencies describing the quantum dynamics in the wells~\cite{kiffner:13l}. 
For the red solid  curve in Fig.~~\ref{fig7}(b) those frequencies are about $10\%$ larger as compared to the  
the case without $np_{1/2}$ states. Note that the parameters in Figs.~~\ref{fig7}(b) and (c) correspond to 
$\Delta/\delta=100$. We find that the impact of the $np_{1/2}$ states on the trimer states 
reported in~\cite{kiffner:13l}  reduces for larger values of $\Delta/\delta$. 
%
%%%%%%%%%%%%%%%%%%%%%%
%%%%%%%%%%%%%%%%%%%%  
%

%
The potential wells indicated by the red solid and blue dashed lines in 
Fig.~~\ref{fig7}(b) are induced by the Stark shift $\delta$. On the contrary, there are 
additional and deeper wells in Fig.~~\ref{fig7}(b) that exist even without an external electric field. 
Since these wells are absent in Fig.~~\ref{fig7}(c), they 
are a consequence of avoided crossings with  potential curves from the $np_{1/2}$ manifold. 
Their physical origin is  thus related to the potential wells shown in Fig.~\ref{fig7}(a). 

So far we analyzed only minima in the potential curves as a function of the
triangle edge length $R$. 
As discussed above the potential minima in Fig.~\ref{fig7}(a) exist even for
$\delta=0$ where the system is spherically symmetric. 
Three-body bound states for $\delta=0$ will thus be invariant under uniform
rotations 
of the relative position vectors $R_{13}$ and $R_{23}$, which is in contrast to
the system described in~\cite{kiffner:13l}. 
A more detailed analysis of the angular dependence of trimer states  
are presented in the next section~\ref{new}  where we investigate potential
curves in the manifold $\mc{E}_1$. 
In contrast to the states in $\mc{E}_2$,  states in $\mc{E}_1$ can have a direct
transition dipole moment with states in $\mc{E}_0$ 
such that they do not need to be excited via a two-photon process. 
\subsection{Potential surfaces in $\mc{E}_1$ \label{new}}
\begin{figure}[t!]
\begin{center}
\includegraphics[width=8.5cm]{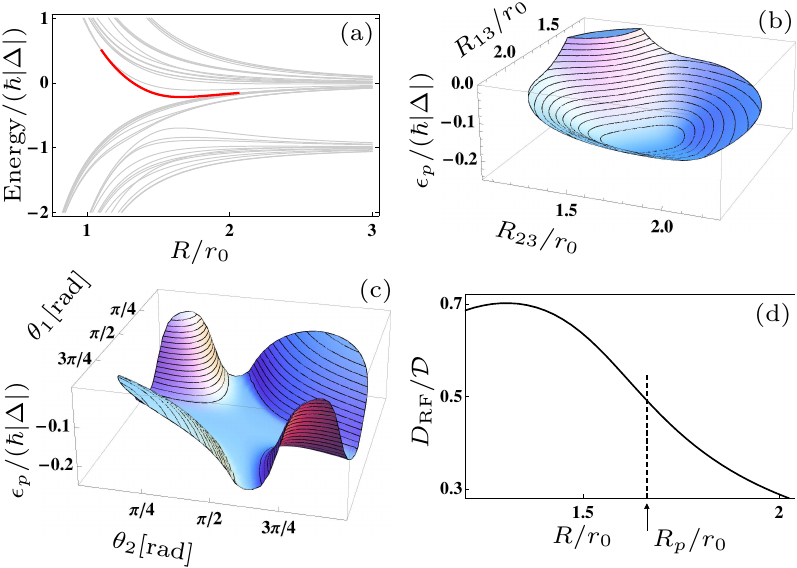}
\end{center}
\caption{\label{fig8}
(Color online) 
(a)  All potential curves within the manifold $\mc{E}_1$. The  red solid  curve
shows $\epsilon_{p}$. 
(b) Variation of the  energy surface
$\epsilon_{p}$ with 
 $R_{13}$ and $R_{23}$ for $\theta_1=\theta_2=\pi/2$, $\phi=\pi/3$ and
$\Delta/\delta=100$. 
(c)  Dependence of $\epsilon_{p}$ on $\theta_1$ and $\theta_2$ for 
$R_{13}=R_{23}=R_{\text{min}}$, $\phi=\pi/3$ and $\Delta/\delta=100$. 
(d) Transition dipole matrix element between the bound state $\ket{\psi_p}$ and
the ground state $\ket{\psi_i}$ in Eq.~(\ref{gstate}) as a function 
of the triangle edge length $R$. 
}
\end{figure}

In this section we investigate the potential curves in the $\mc{E}_1$ manifold
for the geometrical setup shown in Fig.~\ref{fig6}(c). 
All potential curves in $\mc{E}_1$ are shown in Fig.~\ref{fig8}(a) as a function
of the triangle edge length $R$. The red solid  curve 
labels the potential well $\epsilon_{p}$ which  has a local minimum  at
$R_{p} = 1.66 r_0$. 
This minimum corresponds to the parameters $\mf{v}_{p} =
(R_{p},R_{p},\pi/2,\pi/2,\pi/3)$. 
In order to establish that the potential curve $\epsilon_{p}$ has a true
minimum with respect to all independent variables 
we follow the approach outlined in~\cite{kiffner:13l}.  We find that the
gradient of $\epsilon_{p}$ with respect to the 
independent variables $\mf{v}$ vanishes at $\mf{v}_{p}$, and the Hessian
matrix of $\epsilon_{p}$ is positive definite. 
This shows that $\epsilon_{p}$ has indeed a local minimum at
$\mf{v}_{p}$. This result holds only in the presence of the external field since 
the system becomes isotropic for $\delta=0$. 
Note that the equilibrium distance $R_p$ between the atoms in the trimer 
configuration coincides with the atomic separation in the dimer state discussed in Sec.~\ref{diqd}.  
Despite this result the stability of the trimer configuration in the subspace $\mc{E}_1$ 
cannot be fully explained by the pairwise binding energies of the atoms. This would only be the case if the 
trimers  were in a product state. On the contrary, the reduced quantum state of two atoms obtained by tracing out 
one atom is a mixed state and hence the trimer state is entangled. Moreover, 
some  components of the reduced two-atom state reside in the $nsns$ subspace where the atoms are unbound. 

The dependence of $\epsilon_{p}$ on the parameters $\mf{v}$ around the
minimum at $\mf{v}_{p}$ is shown 
in Figs.~\ref{fig8}(b) and~(c). The potential curve $\epsilon_{p}$  has a
deep minimum if $R_{13}$ and $R_{23}$ are varied and the 
remaining parameters are fixed at $\theta_1=\theta_2=\pi/2$ and $\phi=\pi/3$
[see Fig.~\ref{fig8}(b)]. The depth of the potential well is approximately given by $0.1\hbar|\Delta|$ 
corresponding to trapping frequencies of the order of $100\,\text{MHz}$ for $n\approx 40$. 
The potential surface in  Fig.~\ref{fig8}(c) shows $\epsilon_{p}$ as a
function of  $\theta_1$ and $\theta_2$ for $R_{13}=R_{23}=R_{p}$ 
and $\phi=\pi/3$. The four deep valleys in the energy landscape 
correspond to parameters that are energetically equivalent to 
the initial configuration $\mf{v}_{p}$ in the absence of the electric
field.  The small Stark shift $\delta$  breaks the spherical 
symmetry of the system and results in a weak trapping of the trimer  in the plane perpendicular 
to the electric field. 
\subsubsection{Quantum dynamics}
Next we discuss the quantum dynamics of the three Rydberg atoms in the trimer
potential $\epsilon_{p}$ as described in detail in~\cite{kiffner:13l}.  
The frequencies of the normal modes of the trimer configuration are given by
\begin{align}
& \omega_1 = 1.04 \times\omega_{\text{vib}}, &&  \omega_2 =  \omega_3=0.31\times
\omega_{\text{vib}}, \notag \\
& \omega_4 =  \omega_5 = 0.03\times \omega_{\text{vib}}, 
\end{align}
%
%
%%%%%%%%%%%%%%%%%%%%%%
%%%%%%%%%%%%%%%%%%%% 
%
where $\omega_{\text{vib}}$ is defined in Eq.~(\ref{ovib}). 
The frequency $\omega_1$ belongs to the symmetric stretch mode, and the degenerate 
frequencies $\omega_2$ and $\omega_3$ correspond to the  scissor
and asymmetric stretch modes, respectively. The two frequencies $\omega_4$ and  $\omega_5$ 
are equal and describe wagging and twisting, respectively. These frequencies are roughly an order of magnitude smaller 
than the other frequencies because they are only different from zero in the presence of the (small) Stark shift $\delta$. 
Each of the three Rydberg atoms constituting the trimer state is in a coherent superposition of $ns$ and $np$ states. 
The radiative lifetime of each atom can be calculated from the lifetimes of the $ns_{1/2}$ $np_{1/2}$ and 
$np_{3/2}$  Rydberg states~\cite{beterov:09}.  
For Rb atoms with $n=60$ we find that the decay rate  at a temperature of  $300\,\text{K}$  is approximately 
$9\,\text{kHz}$, which is much smaller than the largest oscillation frequency $\omega_1=210\,\text{kHz}$. 
The equilibrium edge length of 
the trimer state for these parameters is $R_p\approx 4.0\,\mu\text{m}$ which can be 
experimentally resolved~\cite{schwarzkopf:11,schauss:12}. 
\subsubsection{Preparation}
The efficient excitation of the trimer state $\ket{\psi_p}$ 
from a state $\ket{\psi_i}$ in the subspace $\mc{E}_0$ via microwave fields requires
a non-zero transition dipole matrix element 
between them. Here we assume that the atoms are initially prepared  in the  state 
\begin{align}
\ket{\psi_i} = \ket{s_{1/2},-1/2,s_{1/2}-1/2,s_{1/2}-1/2}
\label{gstate}
\end{align}
 and consider $\pi$ polarized microwave fields. For the geometry under consideration we find that 
polarization directions perpendicular to $\mf{e}_z$ do not couple $\ket{\psi_i}$ and $\ket{\psi_p}$. 
The excitation Rabi frequency is directly proportional to the dipole
matrix element
\begin{align}
D_{\text{RF}}=
|\bra{\psi_p}\hat{\mathbf{d}}_{3\text{atom}}\cdot\hat{\mf{e}}_z\ket{\psi_i}|,
\label{dthree}
\end{align}
where 
\begin{align}
\mf{\hat{d}}_{3\text{atom}}= \mf{\hat{d}}^{(1)} + \mf{\hat{d}}^{(2)}+
\mf{\hat{d}}^{(3)}.
\end{align}
The value of $D_{\text{RF}}$ as a function of the triangle edge length is shown
in Fig.~\ref{fig8}(d).  Note that we chose the state $\ket{\psi_p}$ in the two-dimensional subspace induced by 
Kramers degeneracy which maximizes $D_{\text{RF}}$. 
Since $D_{\text{RF}}$ is large at the position of the potential
minimum where the Franck-Condon factor 
associated with the transition $\ket{\psi_i}\rightarrow\ket{\psi_{p}}$ is
maximal, efficient excitation of  the trimer state via microwave
radiation is possible. 

\section{Conclusion \label{conclusion}}
In this paper we show that the DD interaction between two and three Rydberg atoms with 
non-overlapping electron clouds can give rise to bound states. 
We focus on two different types of dimer and trimer states. The first one arises from avoided 
crossings between Stark-shifted Rydberg levels that are resonantly coupled by the DD 
interaction. These states were discussed previously in two-atom~\cite{kiffner:12} and 
three-atom~\cite{kiffner:13l} systems in a simplified level scheme ignoring the $np_{1/2}$ 
states in Fig.~\ref{fig1}(a). Here we show that the inclusion of the $np_{1/2}$ states 
leaves the qualitative feature of the dimer  (trimer) states investigated in~\cite{kiffner:12} 
(\cite{kiffner:13l}) unchanged. We provide quantitative corrections for the position and 
oscillation frequencies of the  trimer states discussed in~\cite{kiffner:13l}. 
The second type of bound states arises from avoided crossings between Rydberg states that are separated by 
the energy difference between the $np_{3/2}$ and $np_{1/2}$ manifolds in the  limit of large atomic separations. 
These dimer and trimer states have not been reported previously and do not require an external electric field. 
The depth of the potential wells  and typical 
oscillation frequencies describing the quantum dynamics are enhanced by at 
least one order of magnitude as compared to the states investigated in~\cite{kiffner:12,kiffner:13l}. 
We show that the novel dimer and trimer states can be efficiently excited via microwave fields. Typical 
equilibrium distances of the atoms in the bound states are of the order of several microns, 
and atoms can be prepared and detected in geometries at those length scales~\cite{schauss:12,schwarzkopf:11,beguin:13}. 
We find that electric fields can be employed to align the molecules. The dimer configuration can be aligned 
along the electric field axis and the trimer configuration can be trapped in a plane perpendicular to this axis. 
In conclusion, we are confident that the dimer and trimer states can be produced and detected with existing technology. 

\begin{acknowledgments}
The authors acknowledge financial support from the National Research Foundation
and the Ministry of Education, Singapore. 
\end{acknowledgments}
\appendix

\section{Atomic Hamiltonian and DD Interaction \label{appena}}
The atomic Hamiltonian $H_{\text{A}}^{(\alpha)}$ of atom $\alpha$ 
and with the level scheme depicted in Fig.~\ref{fig1}(a) is given by 
\begin{align}
H_{\text{A}}^{(\alpha)}=\hbar [& \omega_{0} \ket{ p_{3/2}-3/2 }_{\alpha}
\bra{ p_{3/2}-3/2 }_{\alpha} \notag \\
&+\omega_{0} \ket{ p_{3/2}+3/2 }_{\alpha} \bra{ p_{3/2}+3/2}_{\alpha} \notag \\
&+(\omega_{0}+\delta) \ket{p_{3/2}-1/2 }_{\alpha} \bra{ p_{3/2}-1/2 }_{\alpha}
\notag \\
&+(\omega_{0}+\delta)\ket{p_{3/2}+1/2 }_{\alpha} \bra{ p_{3/2}+1/2}_{\alpha}
\notag \\
&+(\omega_{0}+\Delta)\ket{p_{1/2}-1/2 }_{\alpha} \bra{ p_{1/2}-1/2}_{\alpha} \notag \\
&+(\omega_{0}+\Delta)\ket{p_{1/2}+1/2 }_{\alpha} \bra{ p_{1/2}+1/2}_{\alpha}], 
\end{align}
where $\omega_{0}$ is the resonance frequency of the $\ket{ns_{1/2}}
\leftrightarrow \ket{np_{3/2}}$ transition. 

The DD interaction~\cite{tannoudji:api} between atoms $\alpha$ and~$\beta$ located at 
positions $\mf{R}_{\alpha}$ and $\mf{R}_{\beta}$ is defined as 
\begin{align}
V_{\alpha\beta}(\mf{R}) = \frac{1}{4\pi\varepsilon_0 R^3}[\mf{\hat{d}}^{(\alpha)}\cdot\mf{\hat{d}}^{(\beta)}
-3(\mf{\hat{d}}^{(\alpha)}\cdot\vec{\mf{R}})(\mf{\hat{d}}^{(\beta)}\cdot\vec{\mf{R}})], 
\label{vdd}
\end{align}
where $\mf{\hat{d}}^{(\alpha)}$ is the electric dipole-moment operator of atom $\alpha$, 
$\mf{R} = \mf{R}_{\alpha}-\mf{R}_{\beta}$  and 
$\vec{\mf{R}}=\mf{R}/R$ is the corresponding  unit vector. 
Matrix elements of the electric-dipole-moment operator $\mf{\hat{d}}$ of an individual atom 
are evaluated via the Wigner-Eckert theorem~\cite{walker:08,edmonds:amq},  
\begin{eqnarray}
<n\ensuremath{'} l\ensuremath{'}_{
j\ensuremath{'}}m\ensuremath{'}|\hat{d}_{q}|nl_{j}m>
&=&(-1)^{\mathrm{j+l'-1/2}}C_{jm1q}^{j\ensuremath{'}m\ensuremath{'}}\sqrt{2j+1}
\notag \\
&\times & 
\left \lbrace 
\begin{array}{ccc}
l & 1/2 & j \\
j\ensuremath{'} & 1 & l\ensuremath{'} 
\end{array} 
\right \rbrace  
 <n\ensuremath{'}l\ensuremath{'}\parallel \mf{\hat{d}}\parallel nl>, \notag \\
\label{WE}
\end{eqnarray}
where $\hat{d}_{q}$ ($q\in\{-1,0,1\}$) are the spherical components of the
dipole operator, 
\begin{align}
\hat{d}_{1}=-\frac{\hat{d}_{x}+i \hat{d}_{y}}{\sqrt{2}},\quad
\hat{d}_{0}=\hat{d}_{z},\quad
\hat{d}_{-1}=\frac{\hat{d}_{x}-i \hat{d}_{y}}{\sqrt{2}}, 
\label{dq}
\end{align} 
$C_{jm1q}^{j\ensuremath{'}m\ensuremath{'}}$ are Clebsch-Gordan
coefficients, and the $3\times 2$ matrices in curly braces are Wigner $6-j$
symbols.   
The reduced dipole matrix element in~Eq. (\ref{WE}) can be written
as~\cite{walker:08,edmonds:amq} 
\begin{align}
<n\ensuremath{'}l\ensuremath{'}\parallel \mf{\hat{d}}\parallel nl> = \sqrt{2l+1}
C_{l010}^{l\ensuremath{'}0} e \bra{n\ensuremath{'}l\ensuremath{'}}r\ket{nl}, 
\label{red1}
\end{align}
where $e$ is the elementary charge and
$\bra{n\ensuremath{'}l\ensuremath{'}}r\ket{nl}$ is a radial matrix element. 
Combining Eqs.~(\ref{WE}) and~(\ref{red1}) for $n'=n$, $j=1/2$, $l'=p$ and $l=s$, we obtain 
\begin{align}
\bra{np_{3/2}m'}\mf{\hat{d}}\ket{ns_{1/2}m} & =- \mc{D} \sum_{q=-1}^1 
C^{3/2m'}_{1/2 m 1 q} \vec{\mf{\epsilon}}_q, \label{we1}\\
\bra{np_{1/2}m'}\mf{\hat{d}}\ket{ns_{1/2}m} & =  \mc{D} \sum_{q=-1}^1 
C^{1/2m'}_{1/2 m 1 q} \vec{\mf{\epsilon}}_q, \label{we2}
\end{align}
where Eq.~(\ref{we1}) [Eq.~(\ref{we2})] corresponds to $j'=3/2$ ($j'=1/2$). 
The spherical unit vectors $\vec{\mf{\epsilon}}_q$ in Eqs.~(\ref{we1}) and~(\ref{we2}) are defined as 
\begin{eqnarray}
\vec{\mf{\epsilon}}_{1}=-\frac{\vec{\mf{x}}-i \vec{\mf{y}}}{\sqrt{2}},\quad
\vec{\mf{\epsilon}}_{0}=\vec{\mf{z}},\quad
\vec{\mf{\epsilon}}_{-1}=\frac{\vec{\mf{x}}+i \vec{\mf{y}}}{\sqrt{2}}, 
\label{vecs}
\end{eqnarray}
and 
\begin{align}
\mc{D} = \frac{1}{\sqrt{3}}e \bra{n p}r\ket{n s} 
\label{D}
\end{align}
is a reduced dipole matrix element. 
For alkali-metal atoms with $n\ge 40$ we have~\cite{walker:08} 
$\bra{np}r\ket{ns} \approx n^2 a_0$ 
where $a_0$ is the Bohr radius. 
Since the sums in Eqs.~(\ref{we1}) and~(\ref{we2}) are of the order of unity,
the characteristic strength of the DD interaction is given by
\begin{align}
\hbar\Omega = \frac{|\mc{D}|^2}{4\pi\epsilon_0 R^3}. 
\label{char}
\end{align}
The characteristic length scale $r_0$ in Eq.~(\ref{r0}) is obtained by equating
$\Omega$ in Eq.~(\ref{char}) with $|\Delta|$. 

%%%%%%%%%%%%%%%%%%%%%%%%%%%%%%%%%%%%%%%%
%
%Merlin.mbs v4.21 2009-07-09.
%
%
%
\end{document}